\documentclass[a4paper]{article}
\usepackage[latin1]{inputenc}
\usepackage{epic,eepic,xspace,amssymb,amsmath,tensor}

\newcommand{\sfrac}[2]{\textstyle{\frac{#1}{#2}}}

\begin{document}

\title{\bf Inducing the Lovelock action}

\author{Jan-Peter B\"ornsen$^1$\footnote{E-mail: jan-peter.boernsen@desy.de} \ and
		 Anton E.\,M. van de Ven$^2$\footnote{E-mail: A.E.M.vandeVen@uu.nl}\\
\\
         {\small $^1$ II. Institute for Theoretical Physics, University Hamburg, Germany}\\
		 {\small $^2$ University College Utrecht, The Netherlands}
}
\date{\today}

\maketitle

\begin{abstract}
We re-analyze a possible ambiguity in the application of dimensional
regularization to Einstein-Gauss-Bonnet gravity, arising from the way one
treats the Gauss-Bonnet term~\cite{CK79}. It is demonstrated that the
addition of such a term to the action gives rise to a non-minimal graviton
wave operator, but does not produce new on shell divergences at one loop
order in $d=4$. However, from a $d$-dimensional perspective the
Gauss-Bonnet term is shown to generate new divergences in the form of
the six-dimensional Euler density. The conjecture that one would next
produce the eight-dimensional Euler term is shown to be false.
\end{abstract}

\addtocounter{footnote}{2}

\section{Introduction}
The natural generalization of the Einstein-Hilbert action to a 
spacetime of dimension greater than four is provided by the Lovelock 
action~\cite{Lo71,Lo72}; see e.g.~\cite{DM03} for a review. It consists of the sum of all 
dimensionally continued Euler densities, each term coming with its own 
coupling constant. The Lovelock action gives rise to a symmetric
conserved field equation which contains derivatives of the metric no
higher than second order and is quasi-linear in these second
derivatives. This leads to a well-defined classical Cauchy problem~\cite{Ar87,Ch88}, 
and to unitarity at the quantum level, unlike generic
higher derivative actions~\cite{DN74,St76}. Unitarity is also the reason
why one expects the Lovelock action to make its appearance in the
low-energy limit of string theory~\cite{Zw85,DR86}. More recently, numerous 
papers have appeared in which this same action appears in connection
with brane scenarios in string theory, see e.g.~\cite{DM03,Ma03}. This
partly motivated our examination of the ultraviolet behavior of the
Lovelock action.  

It is well known that a perturbative treatment of Einstein gravity
gives rise to a one loop finite S-matrix in four dimensions. The
background field method and dimensional regularization predict one
loop divergences of curvature-squared form~\cite{tHV74}. Although these do not
have the functional form of the classical action, the vacuum field
equations combined with the Gauss-Bonnet identity would imply that
such divergences do not survive. However, this reasoning was
criticized in a little known paper by Capper and Kimber~\cite{CK79}. They
pointed out that dimensional regularization requires one to work
consistently in $d$ dimensions and, although the field equation can
trivially be continued, there does not exist a $d$-dimensional
Gauss-Bonnet identity. In short, the $d\rightarrow 4$ limit of the
expression  
\begin{equation*}
\frac{1}{d-4}\int d^d x \sqrt{g}\,
(R^{\mu\nu\rho\sigma} R_{\mu\nu\rho\sigma}-4R^{\mu\nu}R_{\mu\nu}+R^2)
\end{equation*} 
is ill defined. This motivated Capper and Kimber to add a
Riemann-squared term to the classical action. They then demonstrated
that this extra term has no effect on tree-level graviton-graviton 
scattering. This was shown to be not due to explicit factors of $d-4$, 
as none appear, but rather caused by the imposition of on shell
conditions and setting $d$ equal to four for all external legs. 
Although no one loop calculation was performed, an appeal to
consistency with other regularization schemes led them to conclude
that the one loop S-matrix of standard gravity is finite after 
all. The re-examination in the present paper answers the question
of Capper and Kimber for the divergent parts of the effective 
action at one loop level. No changes are found, so we confirm what one 
would naively have expected from adding such a $d=4$ topological term. 
However, in the process we discover that from the $d$-dimensional
point of view advocated by Capper and Kimber, adding the Gauss-Bonnet
term to the classical action next requires the $d=6$ 
Euler density as a counter term. Hence, it would seem that the 
renormalization process induces the full Lovelock action.

Besides the well known general difficulty of any calculation in 
quantum gravity, a further reason why the problem was not yet
studied beyond tree level is that one is then faced with a non-minimal 
wave operator. If the leading part of a wave operator equals the 
Laplacian, one speaks of a minimal operator. In gauge theories 
this form can usually be arranged by a judicious gauge choice. 
Minimal wave operators allow the use of the powerful Schwinger-DeWitt 
method and convenient background field algorithms are then available. 
We will show that the addition of the Gauss-Bonnet term inevitably
leads to a non-minimal wave operator for the graviton. An elegant 
extension of the Schwinger-DeWitt method to the non-minimal case was
given in~\cite{BV85} but we did not rely on this work. Since the offending
term in the wave operator turns out to be of first order in the
curvature, we may treat it perturbatively and thus return to the minimal 
setting. Using the fully covariant Schwinger-DeWitt method, we
reproduce and extend to curved space an earlier flat space algorithm~\cite{IIMMW85}. 
There, 't Hooft's background field algorithm~\cite{tH73a} was
generalized to second order in the non-minimal part of the wave
operator (see also the recent systematic work of~\cite{AB01,AB02}). 

In related work, Berredo-Peixoto and Shapiro recently demonstrated 
that no new one loop divergences are generated upon adding the 
Gauss-Bonnet term to either conformal~\cite{BS03} or general~\cite{BS04} higher 
derivative gravity. Such theories are known to be renormalizable, 
though not unitary. The authors of~\cite{BS03,BS04} employed dimensional 
regularization and the Schwinger-DeWitt method for their involved
calculations. Interestingly, the Gauss-Bonnet term was shown 
to actually affect the $d$-dimensional renormalization group equations 
and new non-trivial fixed points were found. However, these
papers do not answer the original issue of~\cite{CK79}: Although 
the classical action in~\cite{BS04} includes Einstein-Hilbert and
cosmological terms, a higher derivative gauge fixing term was chosen
which does not allow one to regain the special case of two-derivative 
gravity. Hence, the analysis in~\cite{BS04} excludes the case of Lovelock 
gravity we are interested in.


\section{Perturbative expansion of the Lovelock action}

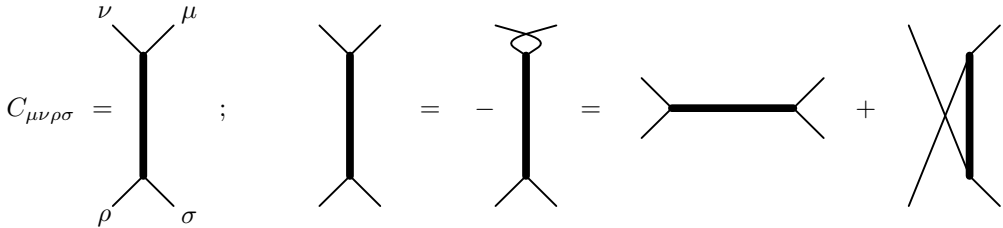
\begin{figure}[h]
\setlength{\unitlength}{2cm}
\begin{picture}(1,2)(-0.6,-0.8)
\put(-0.4,0){$C_{\mu\nu\rho\sigma}\ =$}
\allinethickness{3pt}
\path(0.5,0.4)(0.5,-0.4)
\allinethickness{0.8pt}
\path(0.3,0.6)(0.5,0.4)(0.7,0.6)
\path(0.3,-0.6)(0.5,-0.4)(0.7,-0.6)
\put(0.75,0.65){$\mu$}
\put(0.2,0.65){$\nu$}
\put(0.2,-0.7){$\rho$}
\put(0.75,-0.71){$\sigma$}
\put(1.0,0.0){$;$}
\end{picture}
\begin{picture}(1,2)(-0.9,-0.8)
\allinethickness{3pt}
\path(0.5,0.4)(0.5,-0.4)
\allinethickness{0.8pt}
\path(0.3,0.6)(0.5,0.4)(0.7,0.6)
\path(0.3,-0.6)(0.5,-0.4)(0.7,-0.6)
\end{picture}
\begin{picture}(1,2)(-1.0,-0.8)
\put(-0.2,0){$=\quad -$}
\allinethickness{3pt}
\path(0.5,0.4)(0.5,-0.4)
\allinethickness{0.8pt}
\spline(0.5,0.4)(0.65,0.5)(0.3,0.6)
\spline(0.5,0.4)(0.35,0.5)(0.7,0.6)
\path(0.3,-0.6)(0.5,-0.4)(0.7,-0.6)
\end{picture}
\begin{picture}(1,2)(-1.8,-0.85)
\put(-1,-0.05){$=$}
\allinethickness{3pt}
\path(-0.4,0.0)(0.4,0.0)
\allinethickness{0.8pt}
\path(-0.6,-0.2)(-0.4,0.0)(-0.6,0.2)
\path(0.6,-0.2)(0.4,0.0)(0.6,0.2)
\end{picture}
\begin{picture}(1,2)(-1.8,-0.8)
\put(-0.25,0){$+$}
\allinethickness{3pt}
\path(0.5,0.4)(0.5,-0.4)
\allinethickness{0.8pt}
\path(0.1,-0.6)(0.5,0.4)(0.7,0.6)
\path(0.1,0.6)(0.5,-0.4)(0.7,-0.6)
\end{picture}
\caption{\label{Weyl}Graphical notation for the Weyl-tensor and its symmetries.}
\end{figure}

\begin{figure}[h]
\setlength{\unitlength}{1cm}
\begin{picture}(2,2)(-4.15,-0.8)
\allinethickness{0.8pt}
\put(0.0,0.0){\arc{0.73}{0}{6.28}}
\put(0.0,0.0){\arc{2.1}{0}{6.28}}
\allinethickness{3pt}
\path(-1,0)(-0.4,0)
\path(0.2,0.346)(0.5,0.866)
\path(0.2,-0.346)(0.5,-0.866)
\put(1.2,0){$= C^{3}_W$}
\end{picture}
\begin{picture}(0,0)(-6.65,-0.8)
\allinethickness{0.8pt}
\put(0.0,0.0){\arc{0.73}{1.046}{5.233}}
\put(0.0,0.0){\arc{2.1}{1.046}{5.233}}
\allinethickness{3pt}
\path(-1,0)(-0.4,0)
\path(0.2,0.346)(0.5,0.866)
\path(0.2,-0.346)(0.5,-0.866)
\allinethickness{0.8pt}
\spline(0.183,-0.3161)(0.715,0)(0.909,0.525)(0.675,0,804)(0.602,0.8601)(0.525,0.909) 
\spline(0.183,0.3161)(0.715,0)(0.909,-0.525)(0.675,-0,804)(0.602,-0.8601)(0.525,-0.909) 
\put(1.1,0){$= C^{3}_M$}
\end{picture}
\caption{\label{CWCM}Graphical representation of the Wheel and M\"obius graphs}
\end{figure}
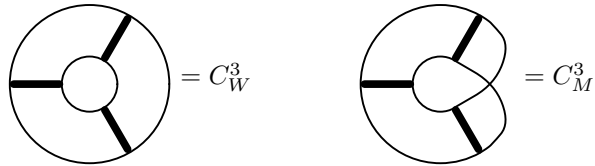

In a $d$ dimensional Riemann space, the Lovelock action is given 
by
\begin{equation}\label{action}
S = \frac{-1}{16\pi G} \sum_{k=0}^{[d/2]} \lambda_k S_k\quad ,\quad 
\lambda_1\equiv 1  \quad ,\quad S_k =\int dv\,\frac{(2k)!}{2^k}\,
R^{[\mu_1\mu_2}_{\,\mu_1\mu_2}\cdots R^{\mu_{2k-1}\mu_{2k}]}_{\,\mu_{2k-1}\mu_{2k}} 
\quad .
\end{equation}
For compactness, the Riemann curvature tensor ${R_{\mu\nu}}^{\rho\sigma}$ 
has been written here as $R_{\mu\nu}^{\rho\sigma}$ and $dv = d^d x\sqrt{g}$.
In $d=2k$, the integrand of $S_k$ is proportional to the Euler density. 
Note the total antisymmetry on $2k$ indices. For given dimension $d$, 
the Schouten identity implies that the maximum value of $k$ is the 
integer part of $d/2$, denoted here as $[d/2]$. The $\lambda_k$ are 
arbitrary coupling constants. In four dimensions this action equals
\begin{eqnarray}
S = \frac{-1}{16\pi G} \int dv\,[\,\lambda_0 + R +\lambda_2
(R^{\mu\nu\rho\sigma} R_{\mu\nu\rho\sigma} -4R^{\mu\nu}R_{\mu\nu} + R^2)\,] 
\end{eqnarray}
One recognizes here the cosmological constant, Einstein-Hilbert term, 
and four-dimensional Euler density, also known as Gauss-Bonnet term. 
The latter is only locally a total derivative and cannot be written 
as the divergence of a vector field. In higher dimensions, further 
dimensionally continued Euler densities appear. For later reference we 
note that in a Ricci flat space $S_3$ reduces to ($C$ represents the
Weyl tensor)  
\begin{equation}\label{S3}
S_3=\int dv\ \tensor{C}{_\mu^\nu_\alpha^\beta} \tensor{C}{_\nu^\rho_\beta^\gamma} 
(\tensor{C}{_\rho^\mu_\gamma^\alpha} -2\tensor{C}{_\rho^\alpha_\gamma^\mu})\,
\equiv\, -\int dv\ (C^3_{\rm W}- 2C^3_{\rm M}) 
\end{equation}
Here, the labels W and M stand for Wheel and M\"obius, respectively.
Using the graphical notation of the Weyl-tensor shown in figure~\ref{Weyl}
one gets a simple graphical representation of $C^3_W$ and $C^3_M$ (see figure~\ref{CWCM})

In order to set the stage for our one loop calculation, we now make
the usual background-quantum splitting via the replacement $g_{\mu\nu}
\rightarrow g_{\mu\nu}+\kappa h_{\mu\nu}$, where $\kappa^2=32\pi G$. 
The action $S$ of Eq~(\ref{action}) then splits into $\sum\kappa^{n-2} S^{(n)}$,
the superscript $n$ indicating the order in the quantum field $h$. 
We find
\begin{equation}\label{action1h}
S^{(1)}= -\int dv\sum_{k=0}^{[(d-1)/2]} 2^{-k}(2k+1)!\lambda_k 
h^{[\mu_1}_{\,\mu_1} R^{\mu_2\mu_3}_{\mu_2\mu_3}\cdots R^{\mu_{2k}\mu_{2k+1}]}_{\mu_{2k}\mu_{2k+1}}
\end{equation}
Note that this expression is totally antisymmetric on $2k+1$ indices, 
so on one more index than in Eq~(\ref{action}). 
The field equation can be read off directly from Eq~(\ref{S3}). E.g. in $d=6$
\begin{equation}\label{fieldeq}
\lambda_0\tensor{\delta}{^\mu_\nu}
+3\tensor{\delta}{^{[\mu}_\nu} \tensor{R}{^{\rho\sigma]}_{\rho\sigma}} 
+30\lambda_2\tensor{\delta}{^{[\mu}_\nu}
\tensor{R}{^{\rho\sigma}_{\rho\sigma}} 
\tensor{R}{^{\tau\omega]}_{\tau\omega}}=0
\end{equation}
Upon writing out the indicated antisymmetrizations, one recognizes
here the Einstein and Bach-Lanczos tensors in the second and third 
term, respectively. Observe that the field equation is covariantly 
conserved: Applying $\nabla_\mu$ and using the Bianchi identity, each 
term vanishes separately, due to the invariance under general
coordinate transformations. The contracted field equation reads
\begin{equation}
d\lambda_0 + (d-2)R +(d-4)\lambda_2 
(R^{\mu\nu\rho\sigma}R_{\mu\nu\rho\sigma}-4R^{\mu\nu}R_{\mu\nu}+R^2) = 0
\end{equation}

At second order in the quantum fields, we find
\begin{eqnarray}
S^{(2)}\!\!\!\! &=&\!\!\!\! \int dv \sum_{k=0}^{[(d-1)/2]} 
2^{-k}(2k+1)!\lambda_k \big[ 
\sfrac12 \big( 
h^{\alpha}_{\mu_1}h^{[\mu_1}_{\,\alpha} - \sfrac12 h h^{[\mu_1}_{\,\mu_1} \big)
R^{\mu_2\mu_3}_{\mu_2\mu_3}\cdots R^{\mu_{2k}\mu_{2k+1}]}_{\mu_{2k}\mu_{2k+1}}  
\nonumber\\
& &\qquad\quad 
+ k\, h^{[\mu_1}_{\,\mu_1} R^{\mu_2\mu_3}_{\mu_2\mu_3}\cdots
                     R^{\mu_{2k-2}\mu_{2k-1}}_{\mu_{2k-2}\mu_{2k-1}}
\big(h^{\mu_{2k};\mu_{2k+1}]}_{\mu_{2k};\mu_{2k+1}} 
+\sfrac12 R^{\mu_{2k}\mu_{2k+1}]}_{\mu_{2k}\ \alpha} h^{\,\alpha}_{\mu_{2k+1}}\big)\big]\label{action2h}
\end{eqnarray}
where we used semi-colon to denote covariant differentiation.
Explicitly in $d=6$
\begin{eqnarray}
\!\!\!\! S^{(2)}\!\!\!\! &=&\!\!\!\! 
\int dv\, \sfrac12\,\big[\lambda_0 (h^\mu_\nu h^\nu_\mu - \sfrac12 h^2)
\nonumber\\
& &\qquad\ \ 
+3\big( ( h^{\alpha}_{\mu}h^{[\mu}_{\alpha} - \sfrac12 h h^{[\mu}_{\mu} ) R^{\nu\rho]}_{\nu\rho}
 +2 h^{[\mu}_{\,\mu} h^{\nu;\rho]}_{\nu;\rho} 
 +  h^{[\mu}_{\,\mu} R^{\nu\rho]}_{\nu\alpha} h^{\alpha}_\rho \big) \label{S2d6} \\
& &\qquad\ \  
+30\lambda_2 \big( (h^{\alpha}_{\mu}h^{[\mu}_{\alpha} 
  - \sfrac12 h h^{[\mu}_{\mu} ) R^{\nu\rho}_{\nu\rho}R^{\sigma\tau]}_{\sigma\tau}
+4 h^{[\mu}_{\,\mu} R^{\nu\rho}_{\nu\rho} h^{\sigma;\tau]}_{\sigma;\tau} 
+2 h^{[\mu}_{\,\mu} R^{\nu\rho}_{\nu\rho} R^{\sigma\tau]}_{\sigma\alpha} h^\alpha_\tau 
\big) \big]
\nonumber
\end{eqnarray}
We emphasize that the form of $S^{(2)}$ is completely determined by 
the requirement of total antisymmetry in its indices; our calculations 
are only needed to find the numerical coefficients in Eq~(\ref{S2d6}). We also 
note that the covariant derivatives in the penultimate term of Eq~(\ref{S2d6})
can be moved via partial integration from one field $h$ to another by 
grace of the Bianchi identity and the indicated total antisymmetry.
The last line of Eq~(\ref{S2d6}) shows that in a general curved background the 
Gauss-Bonnet term does contribute to the quadratic terms and hence to 
the wave operator. Only in special backgrounds, e.g. in a flat space,
does its contribution vanish. 
Eq~(\ref{S2d6}) can be written as $\int h\Delta h$ where the wave operator
takes the schematic form 
\begin{equation}
\Delta\, =\,\sum_{k=1}\lambda_k R^{k-1}\nabla\nabla +\sum_{k=0}\lambda_k R^k
\end{equation} 
Here, $R$ represents the Riemann tensor or any of its contractions. In 
the next section we will show that the leading term of the first sum
can be gauge fixed to the Laplacian but we will also demonstrate 
that the Gauss-Bonnet term unavoidably makes the full wave operator 
non-minimal. In maximally symmetric spaces the wave operator reduces
to the minimal form $(1+c\lambda_2 R)\Box$ with $R$ the Ricci scalar
and $c$ a number. This fact was exploited in~\cite{BD85}. We will not assume 
such a special background.



\section{Gauge fixing the Einstein-Gauss-Bonnet action}

Before we can do quantum calculations, we need to fix the gauge. 
We choose the usual background covariant harmonic, or DeWitt-Feynman, 
gauge. We do so by adding the following gauge fixing term to the
classical action in Eq~(\ref{S2d6})
\begin{equation}\label{Sfix}
S_{fix}=\int dv\ g^{\mu\nu} F_\mu F_\nu \quad , \quad 
F_\mu = h^\nu_{\mu;\nu} - \sfrac12 h_{;\mu} 
\end{equation}
Note that this will gauge fix the Einstein-Hilbert term but not the
other terms. This suffices to define the propagator but does not
affect the contributions coming from the Gauss-Bonnet term. Hence, 
we will wind up with a non-minimal wave operator. One might think that
a more clever gauge choice would return us to a minimal situation. We 
have investigated two options:

\medskip
i) Consider $F_\mu M^{\mu\nu} F_\nu$ with a background dependent but
non-differential operator $M$. This cannot yield a minimal graviton
wave operator because $M$ can only be the Ricci, not the Riemann, tensor. 

\medskip
ii) Consider adding\footnote{Such a gauge was first considered in~\cite{Ic93} 
in the context of Einstein gravity. There, it was erroneously claimed
that the choice of $\zeta$ could affect the numerical coefficient of
the well-known non-renormalizable two loop on shell divergence of this 
theory. See~\cite{LR95} and~\cite{Va95} for the correct statements.} 
$\zeta R_\mu^{\ \,\rho\sigma\nu}h_{\rho\sigma;\nu}$ to $F_\mu$ with new gauge parameter 
$\zeta$. The new terms in $F^2$ then indeed affect the Gauss-Bonnet 
contributions to the quantum action. However, we have verified that no 
choice of the parameter $\zeta$ will cure the wave operator and make
it minimal. 

Thus, we are inevitably led to a non-minimal wave operator. This implies
that at quadratic order in the quantum fields $\phi^ i$, the 
action takes the form
\begin{equation}\label{S2PXW}
S^{(2)} =\int dv\, \big(\sfrac12 P_{ij}\phi^i D^2 \phi^j 
+\sfrac12 X_{ij}\phi^i \phi^j
+\sfrac12 W^{\mu\nu}_{ij} D_\mu\phi^i D_\nu\phi^j \big)
\end{equation}
where $P$, $X$ and $W$ are symmetric matrices which are in general 
background field dependent. Without loss of generality, we may also 
assume that $W^{\mu\nu}=W^{\nu\mu}$. The covariant derivatives satisfy
\begin{equation}
\tensor{[D_\mu,D_\nu]}{^j_{\ k}}\,\phi^k=\tensor{Y}{_{\mu\nu}^j_{\ k}}\, \phi^k 
\end{equation}
The tensor $W$ is dimensionless and the corresponding vertex can be inserted arbitrarily often 
into a one loop diagram without changing its degree of divergence. 
This is the reason why no simple algorithm which generates {\it all} 
one loop divergences for non-minimal operators as in Eq (\ref{S2PXW}) is known.

Some partially successful attempts were made though, on which we now
comment. Pronin and Stepanyantz~\cite{PS97} relied on the diagrammatic
methods of 't Hooft to derive what they call "master formulas 
for the divergent part of the one loop effective action for arbitrary 
(both minimal and non-minimal) operators of any order in 4-dimensional 
curved space".  However, the authors of~\cite{PS97} assume that $W$, called $K$ in 
their work, is covariantly constant (see the remarks following
Eq (28) of~\cite{PS97}). Basically, they assume that the tensor $W$ is
always made from products of metric tensors, but this is not the case 
in the present situation. Avramidi~\cite{Av04} and Avramidi and 
Branson~\cite{AB01,AB02} systematically extended the Schwinger-DeWitt method 
to non-minimal operators, called
non-Laplace type operators there, and diagonal values of the heat
kernel coefficients $a_0$ and $a_1$ (but not yet $a_2$) were 
given\footnote{In~\cite{Av04}, $[a_1]$ is proposed as a possible non-abelian 
generalization of gravity.}. In particular, the so-called 
commutative limit defined in Eq (4.2) of~\cite{Av04} coincides with our approach,
but also in~\cite{AB02} it is assumed that $W$,
called $a$ there, is covariantly constant, which as already noted is
not the case in the present situation. Already in 1985 Barvinsky and Vilkovisky~\cite{BV85}
perturbatively extended the Schwinger-DeWitt method
to non-minimal operators at one loop order. Their main application was
to operators of the kind 
$\delta^\alpha_\beta\Box -\lambda \nabla^\alpha \nabla_\beta$ for
vector fields, where the second term is due to non-minimal gauge
fixing. Non-minimal operators for gravitation were also covered. We
have a similar situation, but the offending contribution generated by the Gauss-Bonnet term is 
already of first order in the background curvature. In the parlance of~\cite{BV85}, our non-minimal
term has background dimensionality $O(1/\ell^2)$. This allows us to treat the $W$-term
as a pertubation which simplifies our analysis a lot\footnote{Cf the analysis in sect. 3 of~\cite{BV85} 
where it is shown that it is better to analyze non-minimal gauge 
theories via the method of Ward identities. This leads namely to
powers of the extremal, i.e. the background field equation.}.
For a flat background spacetime and to second order in $W$, an 
algorithm was first provided Ichinose et al.~\cite{IIMMW85}. This work seems to have gone 
unnoticed so far. No special properties of $W$ were assumed in~\cite{IIMMW85}. 
We have checked and confirmed this algorithm and extended it to a
curved but Ricci flat space plus some terms cubic in $W$. 

Since we will treat the $W$-term perturbatively, we choose to define the propagator 
$G^{kj}(x,x')$ as the inverse of the minimal part of the wave operator
\begin{equation}\label{WO}
( P D^ 2 + X)_{ik} \, G^{kj}\, = -\, \tensor{\delta}{_i^j}\,\delta
\end{equation}
Comparing $S^{(2)}+S_{\rm fix}$, Eq~(\ref{S2d6}) plus (\ref{Sfix}), with Eq~(\ref{S2PXW}), we
read off that for Ricci flat background metric we have
\begin{eqnarray}
P^{\mu\nu\,\rho\sigma} &=&  
g^{\mu(\rho} g^{\sigma)\nu} - \sfrac12 g^{\mu\nu} g^{\rho\sigma}
\nonumber\\
\tensor{X}{^{\mu\nu}_{\rho\sigma}} &=& 
-\lambda_0\delta^{\mu\nu}_{\rho\sigma} + 2\tensor{C}{^\mu_{(\rho}^\nu_{\sigma)}} 
+\lambda_2 (4\tensor{U}{^{\mu\nu}_{(\rho\sigma)}}
-2\tensor{U}{^\mu_{(\rho}^\nu_{\sigma)}} 
-2\tensor{U}{^\mu_{(\rho\sigma)}^\nu} ) 
\label{PXYW}\\
\tensor{(Y_{\kappa\lambda})}{^{\mu\nu}_{\rho\sigma}} &=& 
2\,\tensor{C}{_{\kappa\lambda}^{(\mu}_{(\rho}}
\tensor{\delta}{^{\nu)}_{\sigma)}}  
\nonumber\\
\tensor{(W_{\kappa\lambda})}{^{\mu\nu}_{\rho\sigma}} &=& 
16\lambda_2\tensor{\delta}{^{(\mu}_{(\kappa}} \tensor{C}{_{\lambda)(\rho}^{\nu)}_{\sigma)}}
-8\lambda_2\tensor{C}{_\kappa^{(\mu}_\lambda^{\nu)}} g_{\rho\sigma} 
+8\lambda_2\tensor{\delta}{^{(\mu}_{(\rho}} \tensor{C}{_{\sigma)(\kappa}^{\nu)}_{\lambda)}}
-4\lambda_2 g_{\kappa\lambda} \tensor{C}{^\mu_{(\rho}^\nu_{\sigma)}} 
\nonumber
\end{eqnarray}
Here, the first two terms in $W$ still need to be symmetrized under
pair interchange $\mu\nu\leftrightarrow\rho\sigma$. In fact, $W$ is
then totally symmetric under interchanges $\kappa\lambda\leftrightarrow
\mu\nu\leftrightarrow\rho\sigma$. In the second expression we defined
\begin{equation}
U_{\mu\nu\rho\sigma}\,\equiv\,\tensor{C}{_\mu^\kappa_\rho^\lambda}C_{\nu\kappa\sigma\lambda} 
\end{equation} 
Note that after contracting a pair of Weyl tensors twice with each
other one can always put the indices in this standard order~\cite{Ve92}. 
Tensor $U$ has the following symmetries
\begin{equation}
U_{\mu\nu\rho\sigma} = U_{\rho\sigma\mu\nu} = U_{\nu\mu\sigma\rho}
\end{equation}
Furthermore, the field equation allows us to require $U$ to be 
traceless on any pair of its indices, i.e. we will drop triple or 
fully contracted pairs of Weyl tensors. We note that
\begin{equation}\label{WProperties}
\tensor{(W{_\kappa^{\ \ \kappa}})}{^{\mu\nu}_{\rho\sigma}} = 
\tensor{(W_{\rho\sigma})}{^{\mu\nu}_\kappa^\kappa} = 
4(4-d)\lambda_2 \tensor{C}{^\mu_{(\rho}^\nu_{\sigma)}} \quad , \quad 
\nabla_\kappa
\tensor{(W{^\kappa_{\ \lambda}})}{^{\mu\nu}_{\rho\sigma}} = 0 
\end{equation}
which will be essential in simplifying our analysis. In $d=4$, the 
Gauss-Bonnet term is topological, hence it is invariant under any 
change of the metric and in particular under conformal
transformations. This explains the vanishing of the traces. The
vanishing of the divergence of $W$ has its origin in the general 
coordinate invariance of the Gauss-Bonnet term which was preserved by 
our choice of gauge.

In principle, we should use the configuration-space metric and its 
inverse to raise and lower pairs of indices on the various tensors. However, 
in Ricci flat spaces there is effectively no difference between the 
various forms of $X$ (this assumes we drop three-fold contractions of
two Riemann tensors in the part of $X$ arising from the Gauss-Bonnet term, but such a 
contraction can anyhow be rewritten via the field equation~(\ref{fieldeq}) and does
not contribute to the $C^3$ scalars). The same is true for $Y_{\mu\nu}$. 
In $d=4$ it is also true for $W_{\mu\nu}$ because it is traceless there. 
In particular, the terms in $X$ involving a pair of three-fold 
contracted Weyl tensors can be dropped.


\section{One loop divergences for Einstein-Gauss-Bonnet}
To find the one loop divergences for the Einstein-Gauss-Bonnet action,
we view it as an action of the form presented in Eq (\ref{S2PXW}) and treat
the $W$-vertex perturbatively. To zeroth order in $W$ and in a Ricci flat space,
the divergent part of the one loop effective action for a minimal wave operator
as in Eq (\ref{WO}) is known to be given by 
\begin{eqnarray}
\sfrac12 {\rm Tr}\,\big({\rm ln\ }G\big)_{\rm div} \!\!\! &=& \!\!\!
\frac1{16\pi^2\epsilon}\ {\rm  Tr}\ [a_2]
\nonumber\\
\!\!\! &=& \!\!\! \frac1{16\pi^2\epsilon}\ 
{\rm Tr}\,(\sfrac12 X^2 +\sfrac1{12} Y^2 +\sfrac1{180}C^2)
\end{eqnarray}
Here $\epsilon=4-d$, $[a_2]$ signifies the diagonal value of the second heat kernel 
coefficient~\cite{Fu89}, Tr denotes the functional trace operation, $Y^2=Y^{\mu\nu}Y_{\mu\nu}$
and $C^2=C^{\mu\nu\rho\sigma}C_{\mu\nu\rho\sigma}$. Actually, Eq (\ref{PXYW}) implies that 
only the $X^2$ term contributes to $C^3$ divergences.

To find the first order effect of the $W$-perturbation, we insert one $W$-vertex
in the one loop graph and determine its divergent part in a Ricci flat space, namely

\begin{eqnarray}
\sfrac12 {\rm Tr}\big(W^{\mu\nu} D_\mu G \overleftarrow{D}_{\nu'}\!\big)_{\rm div} 
\!\! &=& \!\!
\frac1{16\pi^2\epsilon}\ {\rm Tr}\,\big( W^{\mu\nu} 
(D_\mu [D_\nu a_1] - [D_\mu D_\nu a_1] +\sfrac12 g_{\mu\nu} [a_2]\big)
\nonumber\\
\!\! &=& \!\!
\frac1{16\pi^2\epsilon}\ {\rm Tr}\,\big(
\sfrac16 W^{\mu\nu}(D_\mu D_\nu X +\tensor{Y}{_\mu^\rho}\tensor{Y}{_{\rho\nu}} -
\sfrac{1}{15} \tensor{C}{_{\mu\rho\sigma\tau}}\tensor{C}{_\nu^{\rho\sigma\tau}}
\nonumber\\
& & \quad\quad\quad
+\,\sfrac14 \tensor{W}{^\mu_\mu} (X^2 +\sfrac13 D^2 X +
\sfrac16 Y^2 +\sfrac1{90}C^2)\big)
\end{eqnarray}
Due to Eq (\ref{WProperties}) and the remarks at the end of section 3 only the
$W^{\mu\nu}\tensor{Y}{_\mu^\rho}\tensor{Y}{_{\rho\nu}}$ term is relevant.

Inserting two $W$-vertices and again finding the divergent part yields
\begin{eqnarray}
& & 
\sfrac14 {\rm Tr}\,
\big( W^{\mu\nu}(x) D_\nu G(x,x') \overleftarrow{D}_{\rho'}
W^{\rho'\sigma'}(x') D_{\sigma'} G(x',x) \overleftarrow{D}_\mu \big)_{\rm div} 
\nonumber\\
& &\quad 
= {\rm Tr}\, \big(
  \sfrac12 [W^{\mu\nu} a_1 W^{\rho'\sigma'} D_{\sigma'} a_0\overleftarrow{D}_\mu] G_{1,\nu\rho'}G_0
  + \sfrac12 [W^{\mu\nu} a_0 W^{\rho'\sigma'} D_{\sigma'} a_1\overleftarrow{D}_\mu] G_{0,\nu\rho'}G_1
\nonumber\\
& &\qquad
+ [W^{\mu\nu} a_1 W^{\rho'\sigma'} a_0\overleftarrow{D}_\mu] G_{1,\nu\rho'}G_{0,\sigma'}
+ [W^{\mu\nu} a_0 W^{\rho'\sigma'} a_1\overleftarrow{D}_\mu] G_{0,\nu\rho'}G_{1,\sigma'}
\nonumber\\
& &\qquad
+\sfrac12 [W^{\mu\nu} a_0 W^{\rho'\sigma'} a_1] G_{0,\nu\rho'}G_{1,\mu\sigma'} 
+\sfrac12 [W^{\mu\nu} a_0 W^{\rho'\sigma'} a_2]G_{0,\nu\rho'}G_{2,\mu\sigma'} 
\nonumber\\
& &\qquad
+ \sfrac14 [W^{\mu\nu} a_1 W^{\rho'\sigma'} a_1] G_{1,\nu\rho'}G_{1,\mu\sigma'} \big)_{\rm div}
\label{WDGWDG} 
\end{eqnarray}
Here we inserted a heat kernel expansion for each Green function, see~\cite{Ve92,JO84I},
and distributed the derivatives, keeping only such 
terms which in the end can contribute to $C^3$. In particular, we
dropped $\tensor{W}{^\mu_\mu}$, $D_\mu \tensor{W}{^\mu_\nu}$, $D_\mu X$ and
$D_\mu D_\nu X$. A list for the divergent products appearing here can
be found in~\cite{Ve92}. We repeat here the few ones that are actually
needed (see Eq (D.19) of~\cite{Ve92})  
\begin{eqnarray}
& & G_{1,\mu\nu'} G_0 = G_{0,\mu\nu'} G_1 = G_{0,\mu} G_{1,\nu} 
=\textstyle{\frac12} g_{\mu\nu}\delta
\nonumber\\
& & G_{0,\mu\nu'} G_{1,\rho}
=\textstyle{\frac16} (g_{\mu\nu}\nabla_\rho - 2g_{\nu\rho}\nabla_\mu - 2g_{\mu\rho}\nabla_\nu)\delta 
\nonumber\\
& & G_{1,\mu\nu'} G_{0,\rho}
=\textstyle{\frac16} (2g_{\mu\nu}\nabla_\rho - g_{\nu\rho}\nabla_\mu  - g_{\mu\rho}\nabla_\nu)\delta
\label{DivGG}\\
& & G_{0,\mu\nu'} G_{2,\rho\sigma'} = 2G_{1,\mu\nu'} G_{1,\rho\sigma'}
=\textstyle{\frac12} g_{(\mu\nu} g_{\rho\sigma)}\delta 
\nonumber\\
& & G_{1,\mu\nu'} G_{0,\rho\sigma'} = 
\sfrac16 C_{\mu(\rho\sigma)\nu}\delta + \nabla\nabla\delta\ {\rm - terms}
\nonumber
\end{eqnarray}
where we omitted a factor $(16\pi^2\epsilon)^{-1}$ on the far right hand
sides. After substituting these expressions, we partially integrate
the covariant derivatives off the $\delta$-functions and perform the 
integration over $x'$. The final result for the relevant terms is
\begin{eqnarray}
\!\!\!\!\!\!\!\! & & \!\!\!\!\!\!\!\! 
\sfrac14 {\rm Tr}\,\big( W^{\mu\nu} D_\nu G\overleftarrow{D}_{\rho'}
W^{\rho'\sigma'} D_{\sigma'} G\overleftarrow{D}_\mu \big)_{\rm div} 
\nonumber\\
& &\qquad
 = \frac1{16\pi^2\epsilon}\,{\rm Tr}\,\big(
 \sfrac1{12} \tensor{W}{^\mu_\nu}   \tensor{W}{^\nu_\mu} X^2 
+\sfrac1{24} \tensor{W}{^\mu_\nu} X \tensor{W}{^\nu_\mu} X 
+\sfrac1{24} \tensor{W}{^\mu_\nu} D^2 \tensor{W}{^\nu_\mu} X 
\\ 
& & \qquad\qquad\qquad\quad
+\sfrac14 \tensor{W}{^\mu_\nu} \tensor{W}{^\nu_\rho} X \tensor{Y}{^\rho_\mu}
+\sfrac1{12} \tensor{W}{^\mu_\nu} X \tensor{W}{^\nu_\rho} \tensor{Y}{^\rho_\mu}
+\sfrac1{12} \tensor{W}{^\mu_\nu} \tensor{W}{^\rho_\sigma} X \tensor{C}{_{\mu\rho}^{\nu\sigma}} 
\big) 
\nonumber
\end{eqnarray}
where the right hand side is local. The last term explicitly involves
the curvature and is absent from the flat-space algorithm presented in~\cite{IIMMW85}. 
We have verified and agree with the complete algorithm of~\cite{IIMMW85};
cf their Eq (15). Note that the sign of $Y$ in the present study is
opposite to that in~\cite{IIMMW85}. 

Finally, because $X$ contains a term of zeroth order in the Weyl tensor, see Eq (\ref{PXYW}), we also 
need the two $W^3 X^2$ invariants, whose coefficients are easily determined.

Adding the various contributions and keeping only terms which can 
produce $C^3$, our result is 
\begin{eqnarray}
\Gamma^{(1)}_{\rm div} &=& \frac1{16\pi^2\epsilon}\, {\rm Tr}
[\sfrac12 X^2 +\sfrac16 \tensor{W}{^\mu_\nu} \tensor{Y}{^\nu_\rho} \tensor{Y}{^\rho_\mu} 
+\sfrac1{12} \tensor{W}{^\mu_\nu} \tensor{W}{^\nu_\mu} X^2
+\sfrac1{24} \tensor{W}{^\mu_\nu} X \tensor{W}{^\nu_\mu} X
\nonumber\\ 
& &
+\sfrac14 \tensor{W}{^\mu_\nu} \tensor{W}{^\nu_\rho} X \tensor{Y}{^\rho_\mu} 
+\sfrac1{12} \tensor{W}{^\mu_\nu} X \tensor{W}{^\nu_\rho} \tensor{Y}{^\rho_\mu} 
+\sfrac1{24} \tensor{W}{^\mu_\nu} D^2 \tensor{W}{^\nu_\mu} X \nonumber\\ 
& & 
+ \sfrac1{12} \tensor{W}{^\mu_\nu} \tensor{W}{^\rho_\sigma} X
\tensor{C}{_{\mu\rho}^{\nu\sigma}} 
+\sfrac1{24} \tensor{W}{^\mu_\nu} \tensor{W}{^\nu_\rho} \tensor{W}{^\rho_\mu} X^2 
+\sfrac1{24} \tensor{W}{^\mu_\nu} \tensor{W}{^\nu_\rho} X \tensor{W}{^\rho_\mu} X 
] \label{G1}
\end{eqnarray}
Calculations with the {\sl Mathematica}~\cite{MMA} package {\sl MathTensor}~\cite{MT} yield 
in the graviton sector
\begin{eqnarray}
{\rm tr}\,[X^2] 
&=& -12\, \lambda_1 \lambda_2 (C^3_{\rm W}-2C^3_{\rm M})\nonumber\\
{\rm tr }\,[\tensor{W}{^\mu_\nu}\tensor{Y}{^\nu_\rho}\tensor{Y}{^\rho_\mu}]
&=& -\ 12\ \lambda_2\ (C^3_{\rm W}-2C^3_{\rm M})\nonumber\\
{\rm tr}\,[\tensor{W}{^\mu_\nu}\tensor{W}{^\nu_\mu} X^2] = 
{\rm tr}\,[\tensor{W}{^\mu_\nu} X \tensor{W}{^\nu_\mu}X]  
&=& 576\, \lambda_0 \lambda_1 \lambda_2^2 (C^3_{\rm W}-2 C^3_{\rm M})\nonumber\\
{\rm tr}\,[\tensor{W}{^\mu_\nu} \tensor{W}{^\nu_\rho} X\tensor{Y}{^\rho_\mu}]= 
{\rm tr}\,[\tensor{W}{^\mu_\nu}X\tensor{W}{^\nu_\rho}  \tensor{Y}{^\rho_\mu}]  
&=& -72\, \lambda_1 \lambda_2^2 (C^3_{\rm W}-2 C^3_{\rm M}) \label{Terms}\\
{\rm tr}\,[\tensor{W}{^\mu_\nu} D^2 \tensor{W}{^\nu_\mu} X]
&=& 0 \nonumber\\
{\rm tr}\,[\tensor{W}{^\mu_\nu} \tensor{W}{^\rho_\sigma} X] \, \tensor{C}{_{\mu\rho}^{\nu\sigma}} 
  &=& 144\, \lambda_0\lambda_2^2 (C^3_{\rm W}-2 C^3_{\rm M})\nonumber\\
{\rm tr}\,[\tensor{W}{^\mu_\nu}\tensor{W}{^\nu_\rho}\tensor{W}{^\rho_\mu} X^2] =
{\rm tr}\,[\tensor{W}{^\mu_\nu}\tensor{W}{^\nu_\rho} X \tensor{W}{^\rho_\mu} X]
&=& 2160\, \lambda_0^2 \lambda_2^3 (C^3_{\rm W}-2 C^3_{\rm M})\nonumber 
\end{eqnarray}
Note that each invariant by itself is proportional to $E_6$. The abbreviations $C^3_W$ and $C^3_M$ were defined in Eq (\ref{S3})
In the here chosen DeWitt-Feynman gauge, the ghost fields do not contribute to the on shell $C^3$ divergences.
Substitution of Eq (\ref{PXYW}) into Eq (\ref{G1}) produces our final answer
\begin{equation}
\Gamma^{(1)}_{\rm div} = -\frac{\lambda_2}{16 \pi^2 \epsilon} \Big[ 2+6\lambda_1 +12\lambda_0\lambda_2 -72 \lambda_0\lambda_1\lambda_2 -180\lambda_0^2\lambda_2^2\Big]
(C^3_{\rm W}-2 C^3_{\rm M})
\end{equation}
This is our main result. It shows that the
addition of the Gauss-Bonnet term to the sum of the Einstein-Hilbert and cosmological terms induces the on shell six-dimensional Euler
density as a new divergence.  

One might conjecture that the $C^4$ divergences would take the form of the dimensionally continued 
eight-dimensional Euler density $E_8$. In a Ricci flat space and using the equation of motion Eq (\ref{fieldeq}), $E_8$ is proportional to 
the expression graphically represented in figure~\ref{E8}.

\begin{figure}[h]
\setlength{\unitlength}{1cm}
\begin{picture}(0,2)(-1,-1)
\allinethickness{3pt}
\path(-0.3,0.6)(0.3,0.6)
\path(-0.3,0.2)(0.3,0.2)
\path(-0.3,-0.2)(0.3,-0.2)
\path(-0.3,-0.6)(0.3,-0.6)
\allinethickness{0.5pt}
\path(-0.3,0.6)(-0.3,0.2)
\path(-0.3,0.2)(-0.3,-0.2)
\path(-0.3,-0.2)(-0.3,-0.6)
\path(0.3,0.6)(0.3,0.2)
\path(0.3,0.2)(0.3,-0.2)
\path(0.3,-0.2)(0.3,-0.6)
\put(0.3,0.0){\arc{1.2}{-1.57}{1.57}}
\put(-0.3,0.0){\arc{1.2}{1.57}{4.71}}
\put(-1.2,-0.1){$2$}
\put(-0.2,-1.1){$C^{4}_W$}
\end{picture}
\begin{picture}(0,0)(-3.2,-1)
\allinethickness{3pt}
\path(-0.3,0.6)(0.3,0.6)
\path(-0.3,0.2)(0.3,0.2)
\path(-0.3,-0.2)(0.3,-0.2)
\path(-0.3,-0.6)(0.3,-0.6)
\allinethickness{0.5pt}
\path(-0.3,0.6)(-0.3,0.2)
\path(-0.3,0.2)(0.3,-0.2)
\path(-0.3,-0.2)(-0.3,-0.6)
\path(0.3,0.6)(0.3,0.2)
\path(0.3,0.2)(-0.3,-0.2)
\path(0.3,-0.2)(0.3,-0.6)
\put(0.3,0.0){\arc{1.2}{-1.57}{1.57}}
\put(-0.3,0.0){\arc{1.2}{1.57}{4.71}}
\put(-1.3,-0.1){$-$}
\put(-0.2,-1.1){$C^{4}_M$}
\end{picture}
\begin{picture}(0,0)(-5.8,-1)
\allinethickness{3pt}
\path(-0.3,0.6)(-0.3,0.2)
\path(-0.3,-0.2)(0.3,-0.2)
\path(0.3,0.6)(0.3,0.2)
\path(-0.3,-0.6)(0.3,-0.6)
\allinethickness{0.5pt}
\path(-0.3,0.6)(0.3,0.6)
\path(-0.3,0.2)(0.3,0.2)
\path(-0.3,0.2)(-0.3,-0.2)
\path(-0.3,-0.2)(-0.3,-0.6)
\path(0.3,0.2)(0.3,-0.2)
\path(0.3,-0.2)(0.3,-0.6)
\put(0.3,0.0){\arc{1.2}{-1.57}{1.57}}
\put(-0.3,0.0){\arc{1.2}{1.57}{4.71}}
\put(-1.5,-0.1){$-2$}
\put(-0.2,-1.1){$C^4_{Pl}$}
\end{picture}
\begin{picture}(0,0)(-8.4,-1)
\allinethickness{3pt}
\path(-0.3,0.6)(-0.3,0.2)
\path(-0.3,-0.2)(0.3,-0.2)
\path(0.3,0.6)(0.3,0.2)
\path(-0.3,-0.6)(0.3,-0.6)
\allinethickness{0.5pt}
\path(-0.3,0.6)(0.3,0.6)
\path(-0.3,0.2)(0.3,0.2)
\path(-0.3,0.2)(0.3,-0.2)
\path(-0.3,-0.2)(-0.3,-0.6)
\path(0.3,0.2)(-0.3,-0.2)
\path(0.3,-0.2)(0.3,-0.6)
\put(0.3,0.0){\arc{1.2}{-1.57}{1.57}}
\put(-0.3,0.0){\arc{1.2}{1.57}{4.71}}
\put(-1.5,-0.1){$-4$}
\put(-0.2,-1.1){$C^4_{K1}$}
\end{picture}
\begin{picture}(0,0)(-11,-1)
\allinethickness{3pt}
\path(-0.3,0.6)(-0.3,0.2)
\path(0.3,-0.2)(0.3,-0.6)
\path(-0.3,-0.2)(-0.3,-0.6)
\path(0.3,0.6)(0.3,0.2)
\allinethickness{0.5pt}
\path(-0.3,-0.6)(0.3,-0.6)
\path(-0.3,-0.2)(0.3,-0.2)
\path(-0.3,0.6)(0.3,0.6)
\path(-0.3,0.2)(0.3,0.2)
\path(-0.3,0.2)(0.3,-0.2)
\path(0.3,0.2)(-0.3,-0.2)
\put(0.3,0.0){\arc{1.2}{-1.57}{1.57}}
\put(-0.3,0.0){\arc{1.2}{1.57}{4.71}}
\put(-1.5,-0.1){$+4$}
\put(-0.2,-1.1){$C^4_{K2}$}
\end{picture}
\caption{\label{E8}Graphical representation of the Euler density $E_8$}
\end{figure}
\noindent
If we limit our attention to the $\lambda_1^2\lambda_2^2$ sector
we find that only the third and fourth terms in Eq (\ref{G1}) contribute.
These two terms yield a divergence proportional to 

\begin{equation}
6 C_W^4 + 6 C_M^4 - 17 C_{Pl}^4 - 7 C_{K1}^4 + 10 C_{K2}^4 \quad . 
\end{equation}
This shows that, in contrast to the $E_6$ results, the $C^4$ divergences in the
$\lambda_1^2\lambda_2^2$ sector do not take the form of the Euler density $E_8$.

\section{Conclusions and discussion}

We have demonstrated that the addition of a Gauss-Bonnet term to an action consisting of the Einstein-Hilbert term plus cosmological term induces new on shell divergences at one loop order of the form of the six-dimensional Euler density $E_6$. From a strict four-dimensional point of view this implies that the Gauss-Bonnet term has no influence on the one loop renormalizability of gravity. We thus lay to rest this issue, raised long ago by Capper and Kimber~\cite{CK79}, at least for the divergent parts of one loop diagrams. The question remains open for finite one loop scattering processes and also for a possible influence of the Gauss-Bonnet term on the two loop divergences of gravity. Still, given the topological nature of the Gauss-Bonnet term in $d=4$, our result is as expected. 
{\sl Ex nihilo nihil fit.}

In retrospect we might say that 't Hooft and Veltman~\cite{tHV74} showed that if one starts from the Einstein-Hilbert term one induces the Gauss-Bonnet term at one loop. Later, Christensen and Duff~\cite{CD80} added a cosmological constant to the classical action with again the Gauss-Bonnet term being induced. Of course, power counting suffices to predict this, because there is only a single scalar quadratic in the Weyl tensor. This is not anymore so if one includes the Gauss-Bonnet term in the classical action as we did.  There then exist two different cubic scalars, but our calculations show that they appear only in the combination corresponding to the six-dimensional Euler density. It leads us to the conclusion that from the $d$-dimensional point of view advocated by Capper and Kimber~\cite{CK79}, one loop renormalizability of Einstein gravity requires one to extend it to a Lovelock theory. 
However, preliminary calculations show that at quartic order in the Weyl tensor it is not the eight-dimensional Euler density $E_8$ which appears, but rather several non-topological invariants which vanish in $d=4$. It is conceivable that one can arrange for such extra terms to produce just $E_8$ by a special choice of coupling constants. That would constitute an interesting constraint on the parameters of the class of all Lovelock theories. So far, such constraints were based on stability~\cite{Is86,BD85} or causality~\cite{GK06}; see also~\cite{Pr06}. We hope to return to this issue in future. 

Our work can also be seen as an interesting setting in which non-minimal wave operators occur. We have verified the one loop algorithm of~\cite{IIMMW85} for such operators and extended it to curved Ricci flat spaces, both via 't Hooft's non-covariant method and by using the covariant Schwinger-DeWitt method. Our result does not depend on the precise coefficients in the algorithm: Every contribution individually reduces to the six-dimensional Euler density. 

Possibly our work can be continued to all orders in the non-minimal term by extending the methods of~\cite{BV85,AB01,AB02,PS97} to the case of divergenceless rather than covariantly constant $W$-tensor.

\section*{Acknowledgements}
Jan-Peter Börnsen is grateful for financial support of the German Research Foundation (DFG) BO1968/1.
Anton van de Ven appreciates the support of University College Utrecht.

\bibliographystyle{h-physrev3}
\bibliography{ref}

\begin{thebibliography}{10}

\bibitem{CK79}
D.~M. Capper and D.~Kimber,
\newblock J. Phys. {\bf A13}, 3671 (1980).

\bibitem{Lo71}
D.~Lovelock,
\newblock J. Math. Phys. {\bf 12}, 498 (1971),
\newblock idem, Aequationes Math. 4, 127 (1970).

\bibitem{Lo72}
D.~Lovelock,
\newblock J. Math. Phys. {\bf 13}, 874 (1972).

\bibitem{DM03}
N.~Deruelle and J.~Madore,
\newblock {\it On the quasi-linearity of the Einstein-'Gauss-Bonnet' gravity
  field equations},
\newblock in {\em {\rm Proceedings of the Conference in Honor of Yvonne
  Choquet-Bruhat, Cargese}}, 2002, gr-qc/0305004.

\bibitem{Ar87}
C.~Aragone,
\newblock Phys. Lett. {\bf B186}, 151 (1987),
\newblock idem, {\it Stringy Characteristics Of Effective Gravity} in SILARG 6,
  Proceedings, M. Novello (ed.), World Scientific, 1987.

\bibitem{Ch88}
Y.~Choquet-Bruhat,
\newblock J. Math. Phys. {\bf 29}, 1891 (1988).

\bibitem{DN74}
S.~Deser and P.~van Nieuwenhuizen,
\newblock Phys. Rev. {\bf D10}, 401 (1974).

\bibitem{St76}
K.~S. Stelle,
\newblock Phys. Rev. {\bf D16}, 953 (1977).

\bibitem{Zw85}
B.~Zwiebach,
\newblock Phys. Lett. {\bf B156}, 315 (1985),
\newblock idem, {\it Ghost Free $R^2$ Action Arising In String Theory} in
  Argonne 1985, Proceedings, Anomalies, Geometry, Topology, 373.

\bibitem{DR86}
S.~Deser and A.~N. Redlich,
\newblock Phys. Lett. {\bf B176}, 350 (1986),
\newblock Erratum ibid. {\bf B186}, 461 (1987).

\bibitem{Ma03}
R.~Maartens,
\newblock Living Rev. Rel. {\bf 7}, 7 (2004), gr-qc/0312059.

\bibitem{tHV74}
G.~'t~Hooft and M.~J. Veltman,
\newblock Annales de l'institut H.~Poincaré (A) Phys.\ theor. {\bf 20}, 69
  (1974).

\bibitem{BV85}
A.~O. Barvinsky and G.~A. Vilkovisky,
\newblock Phys.\ Rept. {\bf 119}, 1 (1985).

\bibitem{IIMMW85}
S.~Ichinose, C.~Itoh, T.~Minamikawa, K.~Miura, and T.~Watanabe,
\newblock Nucl. Phys. {\bf B260}, 113 (1985).

\bibitem{tH73a}
G.~'t~Hooft,
\newblock Nucl. Phys. {\bf B62}, 444 (1973).

\bibitem{AB01}
I.~Avramidi and T.~Branson,
\newblock Rev. Math. Phys. {\bf 13}, 847 (2001).

\bibitem{AB02}
I.~Avramidi and T.~Branson,
\newblock J. Funct. Anal. {\bf 190}, 292 (2002), hep-th/0109181.

\bibitem{BS03}
G.~de~Berredo-Peixoto and I.~L. Shapiro,
\newblock Phys. Rev. {\bf D70}, 044024 (2004), hep-th/0307030.

\bibitem{BS04}
G.~de~Berredo-Peixoto and I.~L. Shapiro,
\newblock Phys. Rev. {\bf D71}, 064005 (2005), hep-th/0412249.

\bibitem{BD85}
D.~G. Boulware and S.~Deser,
\newblock Phys. Rev. Lett. {\bf 55}, 2656 (1985).

\bibitem{Ic93}
S.~Ichinose,
\newblock Nucl. Phys. {\bf B395}, 433 (1993).

\bibitem{LR95}
P.~M. Lavrov and A.~A. Reshetnyak,
\newblock Phys. Lett. {\bf B351}, 105 (1995), hep-th/9503195.

\bibitem{Va95}
D.~V. Vassilevich,
\newblock Nucl. Phys. {\bf B454}, 685 (1995), hep-th/9509069.

\bibitem{PS97}
P.~I. Pronin and K.~V. Stepanyantz,
\newblock Nucl. Phys. {\bf B485}, 517 (1997), hep-th/9605206.

\bibitem{Av04}
I.~G. Avramidi,
\newblock JHEP {\bf 07}, 030 (2004), hep-th/0406026.

\bibitem{Ve92}
A.~E.~M. van~de Ven,
\newblock Nucl. Phys. {\bf B378}, 309 (1992).

\bibitem{Fu89}
S.~A. Fulling,
\newblock {\em Aspects of Quantum Field Theory in curved space-time} (Cambridge
  University Press, 1989).

\bibitem{JO84I}
I.~Jack and H.~Osborn,
\newblock Nucl. Phys. {\bf B234}, 331 (1984).

\bibitem{MMA}
S.~Wolfram,
\newblock {\em The Mathematica Book}, 4th ed. (Wolfram Media/Cambridge
  University Press, 1999).

\bibitem{MT}
L.~Parker and S.~M. Christensen,
\newblock {\em MathTensor: A System for Doing Tensor Analysis by Computer}
  (Addison-Wesley, 1994).

\bibitem{CD80}
S.~M. Christensen and M.~J. Duff,
\newblock Nucl. Phys. {\bf B170}, 480 (1980).

\bibitem{Is86}
K.~Ishikawa,
\newblock Phys. Lett. {\bf B188}, 186 (1987).

\bibitem{GK06}
A.~Gruzinov and M.~Kleban,
\newblock (2006), hep-th/0612015.

\bibitem{Pr06}
P.~Prester,
\newblock JHEP {\bf 02}, 039 (2006), hep-th/0511306.

\end{thebibliography}

\end{document}